# In-plane anisotropic optical and mechanical properties of two-dimensional MoO$_3$

Sergio Puebla [1], Roberto D'Agosta [2,3], Gabriel Sanchez-Santolino[4], Riccardo Frisenda [1], Carmen Munuera [1] and Andres Castellanos-Gomez [1,*]

[1] Materials Science Factory. Instituto de Ciencia de Materiales de Madrid (ICMM-CSIC), Madrid, E-28049, Spain.

[2] Nano-Bio Spectroscopy Group and European Theoretical Spectroscopy Facility (ETSF), Departamento Polímeros y Materiales Avanzados: Física, Química y Tecnología, Universidad del País Vasco UPV/EHU, Avenida Tolosa 72, E-20018 San Sebastián, Spain.

[3] IKERBASQUE, Basque Foundation for Science, Plaza Euskadi 5, E-48009 Bilbao, Spain

[4] GFMC, Departamento de Física de Materiales & Instituto Pluridisciplinar, Universidad Complutense de Madrid, 28040 Madrid, Spain

* Correspondence: andres.castellanos@csic.es

**Abstract:** Molybdenum trioxide (MoO$_3$) in-plane anisotropy has increasingly attracted the attention of the scientific community in the last few years. Many of the observed in-plane anisotropic properties stem from the anisotropic refractive index and elastic constants of the material but a comprehensive analysis of these fundamental properties is still lacking. Here we employ Raman and micro-reflectance measurements, using polarized light, to determine the angular dependence of the refractive index of thin MoO$_3$ flakes and we study the directional dependence of the MoO$_3$ Young's modulus using the buckling metrology method. We found that MoO$_3$ displays one of the largest in-plane anisotropic mechanical properties reported for 2D materials so far.

**Introduction**

Recently, two-dimensional (2D) materials with an in-plane anisotropy, such as several transition metal chalcogenides, group-VA, black phosphorus and compounds made of two group-VA elements (so called V-V binary materials), have been extensively studied[1–21]. Some of them have demonstrated a great potential in optoelectronics and flexible electronics applications[22–25], allowing the fabrication of devices with new functionalities (e.g. polarization sensitive photodetectors[22,26]) and the observation of novel quasi one-dimensional physics phenomena[27]. Among the available anisotropic 2D materials, molybdenum trioxide is a wide band gap semiconductor ($>2.7eV$)[28], which makes it quasi-transparent in the visible spectrum while being still electrically conductive. This nanomaterial has been proven useful for gas sensing[29], resistive memory technology[30], optoelectronic[31–40], electrochromic[41] and flexible[42,43] applications. The difference between the in-plane (*a-c*) lattice parameters makes α-MoO$_3$ a perfect candidate to investigate optical, mechanical, and electrical in-plane anisotropic properties. Indeed, calculations show that in-plane carrier mobility exhibit strong anisotropic behavior[44] and highly anisotropic propagation of phonon polaritons have been recently observed in α-MoO$_3$ [45–48]. In comparison with metallic plasmon polaritons, phonon polaritons can achieve reduced optical losses, improved light confinements and higher quality factors[49–52]. These interesting optical properties are mainly caused by the anisotropy of the fundamental properties of molybdenum trioxide, which until now have been scarcely investigated.

Here, using a combination of experiments and density functional theory calculations, we have studied the direction-dependent refractive index (birefringence) and Young's Modulus of α-MoO$_3$ exfoliated flakes, two fundamental quantities that govern the anisotropy observed in Raman and phonon polariton experiments. We studied thin flakes of α-MoO$_3$ by aberration corrected scanning transmission electron microscopy and energy-dispersive x-ray spectroscopy. We then deposited the flakes on SiO$_2$/Si substrates and identified the crystallographic orientation of samples using angle-resolved polarized Raman spectroscopy technique[53]. Then, we determined the in-plane anisotropy of the $MoO_3$ refractive index using angle-resolved polarized micro-reflectance spectroscopy, finding a remarkably large birefringence. The anisotropic mechanical properties of the $MoO_3$ flakes was experimentally investigated with the buckling metrology method[54,55], finding one of the largest Young's modulus anisotropy reported so far for 2D materials.

**Results and Discussion**

We have grown $MoO_3$ flakes in atmospheric conditions using a modified version of the hot plate-based physical vapor transport method described in Ref. [56]. Briefly, a molybdenum foil was oxidized on a hotplate at 540°C, then a silicon wafer was placed on top. At this temperature, the molybdenum oxide sublimes and re-crystallizes on the surface of the Si wafer, which is at a slightly lower temperature, forming $MoO_3$ flakes. Then, a Gel-Film (Gel-Pak WF x4 6.0 mil) stamp is used to pick-up and exfoliate the $MoO_3$ flakes. These flakes can be then transferred onto a target substrate (e.g., a holey $Si_3N_4$ TEM grid or a 297 nm $SiO_2$/Si substrate) using deterministic transfer set up (see Materials and Methods section for more information).

Figure 1a shows the layered crystal structure of α-$MoO_3$[57–59], belonging to the *Pbnm* space group. It consists of a double-layer stacking of linked distorted $MoO_6$ octahedra in the *b* direction, along which the adjacent layers are linked by weak van der Waals forces, while in-plane atoms are strongly bonded. This configuration leads to lattice parameters: $a$ = 3.96 Å, $b$ = 13.86 Å and $c$ = 3.70 Å (JCPDS file: 05-0508)[44,53,56,59,60]. We characterized the structure and composition of the grown $MoO_3$ flakes using scanning transmission electron microscopy (STEM) along with energy dispersive x-ray spectroscopy (EDS) and scanning electron microscopy (see Figure S4 in the Supporting Information). A thin flake was transferred into a porous $Si_3N_4$ membrane, as shown in Figure 1b. Atomic resolution high angle annular dark field (HAADF) imaging shows an orthorhombic α-MoO$_3$ structure with a difference in the *a-c* lattice parameters, as depicted also in a selected area diffraction pattern (SAED) acquired at the same flake (Figure 1 c, d).[44,53,56,59,60] The chemical composition of the flake was determined by EDS, Figure 1e shows the spectrum of MoO$_3$ used for quantification, where we find a small oxygen deficiency.

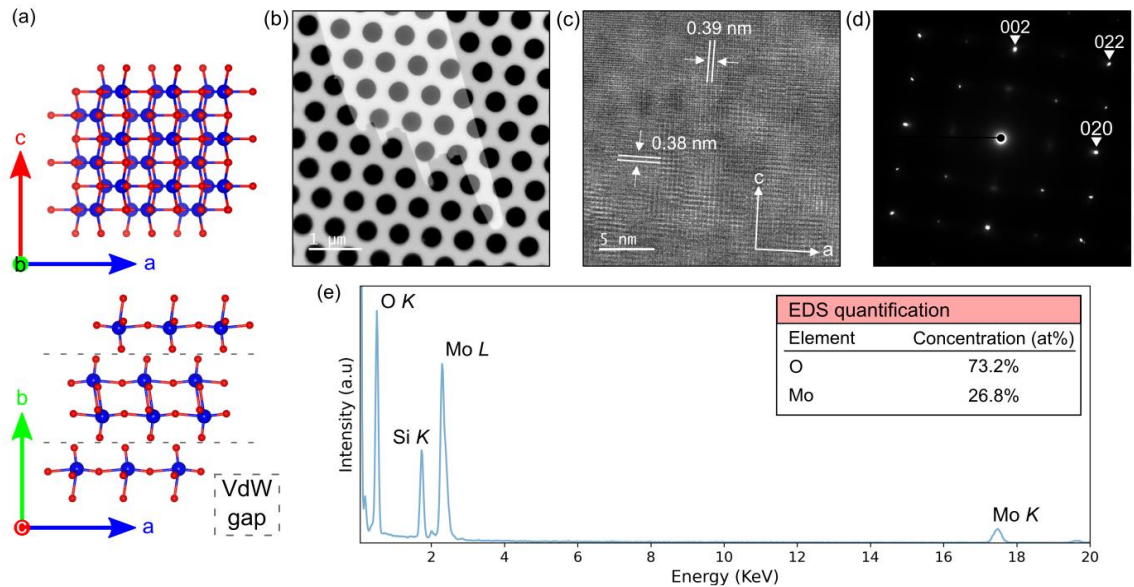

**Fig. 1** (a) Crystal structure of MoO$_3$ in Pbnm notation, spheres in blue (red) represent Mo (O) atoms, the two different views belong to ca (top), and ba plane projection (bottom). (b) Low magnification HAADF image of a mechanically exfoliated MoO$_3$ flake transferred over a porous SiN membrane. (c) Atomic resolution HAADF image depicting the anisotropy between the in-plane (a-c) lattice parameters. (d) SAED pattern characteristic of the orthorhombic α- MoO$_3$. (e) EDS spectrum taken at the flake shown in (b).We have used angle-resolved polarized Raman spectroscopy technique to identify the different crystal directions in our MoO$_3$ flakes. Using a linearly polarized laser in a Raman system (see Materials and Methods section) and setting the analyzer and polarizer in a parallel configuration, while we rotate the sample, we obtain the spectra shown in Fig. 2a. The system set up is depicted in Fig. S1 in the Supplementary Information and it is explained in detail by Liu, X. et. al[61]. Typical MoO$_3$ Raman modes are A$_g$ and B$_g$[62] and no correlation has been found between the Raman modes and the MoO$_3$ thickness[63]. We highlight the peaks centered at 282 cm$^{-1}$, assigned to B$_{2g}$ mode, and at 156 and 818 cm$^{-1}$, assigned to A$_g^c$ and A$_g^a$ mode, respectively. The A$_g^c$ peak corresponds to the translation vibration of the rigid $MoO_6$ octahedra chains along the *c*-axis, and the $A_g^a$ mode is the asymmetric stretching vibration of O-Mo-O atoms along the *a*-axis[53,62].

We have carried out angle-resolved polarized Raman measurements in a $MoO_3$ flake of 28 nm of thickness from 0º to 360º, with a step of 4º. The thickness has been determined by combination of atomic force microscopy with recently developed optical microscopy based techniques[64]. In Fig. 2a three of these measurements at different angles (0º, 45º and 90º) are displayed. Notice the difference in intensity of each mode, revealing how the relative orientation between the polarization angle and the direction of the crystalline axes plays an important role in the Raman process. In fact, previous works have shown how the $A_g$ and $B_{2g}$ mode intensities can be fitted to: [53,62]

$$I(A_g) \propto (A\cos^2\beta + C\sin^2\beta)^2 \quad (1)$$

$$I(B_{2g}) \propto E^2 \sin^2 2\beta \quad (2)$$

where β is the relative angle between the *a*-axis crystal direction and the linear polarization direction of the laser. Figure 2b shows the normalized intensity angle-resolved polarized Raman measurements of each mode (light red), with the resulting fit to equations (1) and (2) (in dark red), showing an excellent agreement. Note, the $A_g^c$ and $A_g^a$ modes are useful to extract the crystal structure of the sample due to their 180-degree periodicity. The inset in Figure 2a shows a microscopy image of the $MoO_3$ flake on $SiO_2$/Si, studied with the Raman spectroscopy. The *c*- and *a*-axes, determined from the Raman measurements are depicted in the figure and coincide with the straight edges produced during the exfoliation of the $MoO_3$ flake.

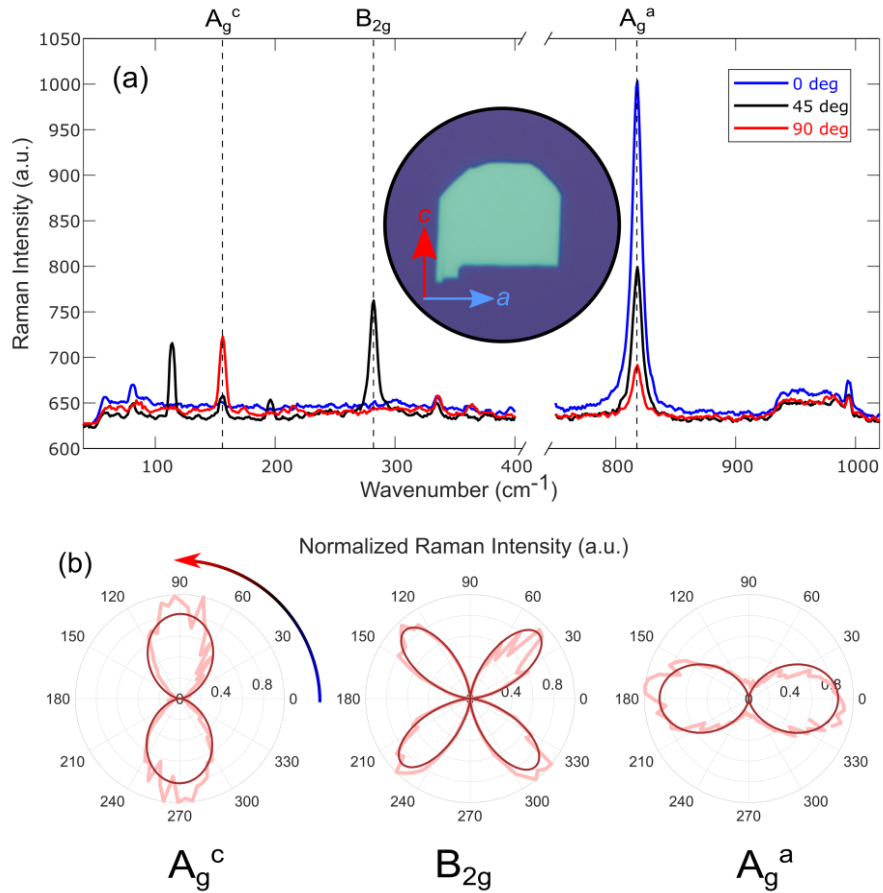

**Fig. 2** (a) Raman spectra at 0º (blue), 45º (black) and 90º (red) angles with respect to the horizontal axis. $A_g^c$, $B_{2g}$ and $A_g^a$ Raman modes are highlighted with vertical dashed lines. Inset: Microscopy image of the sample placed in the initial position (0º) and the *c* and *a* axes, determined from the angle-resolved Raman measurements, are shown. (b) Polar plots, in light red; and fittings, in dark red, of angle-resolved normalized Raman intensities of the $A_g^c$ (left), $B_{2g}$ (middle) and $A_g^a$ (right) modes.

In order to gain an insight about the in-plane anisotropic optical properties of $MoO_3$ flakes we carried out micro-reflectance measurements employing linearly polarized incident light. Figure 3 shows the optical contrast spectra acquired for different alignment between the crystal axis and the incident linear polarization (see the bottom inset). The optical contrast *C* is defined as:

$$C = \frac{I_{fl} - I_{sub}}{I_{fl} + I_{sub}} \quad (3)$$

where $I_{fl}$ and $I_{sub}$ are the intensity measured on the MoO$_3$ flake and on the bare substrate, respectively. Interestingly, it has been demonstrated how one can use a simple Fresnel law-based model to accurately reproduce the measured optical contrast spectra[64]. Moreover, given the known refractive indexes of air, SiO$_2$ and Si and the known thickness of the SiO$_2$ film and MoO$_3$ flake one can determine the refractive index of the MoO$_3$ flake for different alignment between the incident linearly polarized light and the crystal directions by using the refractive index as a fitting parameter to achieve the best fit of the Fresnel law-based model to the experimental data. The top inset in Figure 3 shows the resulting in-plane angular dependence of the refractive index displaying a marked anisotropy (birefringence). While along the a-axis the refractive index of MoO$_3$ is $n_a = 2.21 + 0i$, along the c-axis the refractive index increases up to $n_c = 2.30 + 0i$. The difference $\Delta n$ between $n_c$ and $n_a$, is ~0.1. This value is comparable to that of well-known strongly birefringent materials like calcite (-0.17) and barium borate (-0.12).[65,66] If we compare it with other anisotropic 2D materials, the birefringence of MoO$_3$ is larger than that of ReS$_2$ and ReSe$_2$ (~0.04)[6] but substantially lower than the values reported for black phosphorus (0.25)[6] or TiS$_3$ (0.3)[67]. However, note that, unlike MoO$_3$, all these anisotropic 2D materials are rather opaque in the visible range of the electromagnetic spectrum, limiting their application in polarization optics applications. Therefore, the birefringence value of MoO$_3$, although more modest in comparison with TiS$_3$ or black phosphorus, can have a stronger impact in future ultrathin polarization optics applications. These experimental results for MoO$_3$ are confirmed by theoretical ab-initio calculations through the solution of the Bethe-Salpeter's equation, in which we have obtained refractive index values of $n_a$=2.40 and $n_c$=2.60 at small frequency, with a difference $\Delta n$ ~0.2. We find these theoretical values close to the ones obtained experimentally. We attribute the lower birefringence experimental value to the presence of defects in the $MoO_3$ flakes, not considered in the calculations, that can effectively reduce the anisotropy of the lattice.

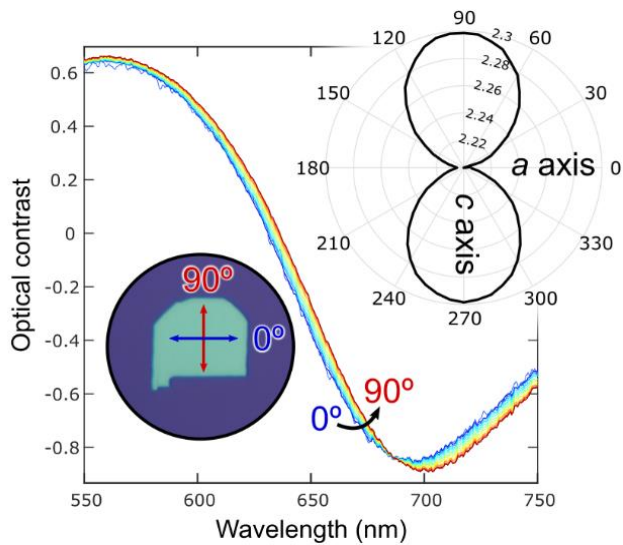

**Fig. 3** Optical contrast *vs*. wavelength as function of the angle of the flake from 0º to 90º of the $MoO_3$ of a 28 nm of thickness flake shown, which is horizontally oriented along the *a*-axis. Inset: Polar plot of the change in refractive index as a function of the relative orientation between the incident linearly polarized light and the *a*-axis of the crystal.

In the following we focus on the characterization of the anisotropy of the Young's modulus, one of the fundamental magnitudes that govern the mechanical properties of materials, of $MoO_3$ flakes. We use buckling induced metrology method, which has been proved to be an easy, but reliable way to study the Young's modulus of thin films[54,55] and 2D materials[68–72]. The method relies on studying the buckling instability that arises when a thin film is deposited onto an adhesive compliant substrate, and it is subjected to in-plane uniaxial compression[73]. Because of this compression, there is a trade-off in the energy related to the adhesion forces between film and substrate and the bending rigidity of the film. This trade-off leads to a rippling pattern of the thin film, which is characteristic of the film and substrate properties (in the Supplementary Information can be found an animated GIF, Supplementary GIF, of a α-MoO$_3$ flake compressed in this way). To perform the buckling metrology measurements, the MoO$_3$ flakes were transferred onto a flat (unstrained) Gel-Film substrate that is mounted on a rotation stage under the inspection of an optical microscope. Then the flakes were subjected to compressive strain along different crystal directions by pinching the surface of the Gel-Film with two glass slides and pictures of the obtained ripple patterns are acquired with a digital camera attached to the microscope. The MoO$_3$ flakes are then transferred to a SiO$_2$/Si substrate to determine their crystal orientation, through Raman spectroscopy and micro-reflectance, and thickness through AFM (see Figure S2 and S3 of the Supporting Information).

The wavelength, λ, of the rippling pattern can be used to determine the Young's modulus:

$$E_f = 3 \left(\frac{\lambda}{2\pi h}\right)^3 \frac{1 - \nu_{fa} \nu_{fc}}{1 - \nu_s^2} E_s \qquad (4)$$

being $h$ the flake thickness, $\nu_s$, $\nu_f$, $E_s$ and $E_f$ the Poisson's ratio and Young's Modulus of substrate and flake, respectively. The Poisson's ratio of α-MoO$_3$ must be taken along the two axes. Using density functional theory (more details available in the Supplementary Information) we calculate Poisson's ratios of α-MoO$_3$ correspond to $\nu_{fa}$=0.147 and $\nu_{fc}$=0.06. In our experiment we used a Gel-Film substrate as compliant substrate, with $\nu_{sub}$=0.5 [74] and its Young's Modulus is $E_{sub}$=492 ± 11 kPa [68].

Figure 4a shows optical images of a 24 nm thickness MoO$_3$ flake (thickness and orientation obtained with optical microscopy based technique, shown in Figure S2) when it is subjected to compression along the c- and a-axes. Upon compression along different orientations the MoO$_3$ develops a rippled pattern whose periodicity depends on the direction. In Fig. 4b, a statistical study is shown with 14 different samples with thicknesses ranging from 18 to 62 nm (white circles), from which we obtain the mean Young's Modulus values and their standard deviation along the a- and c-axis. We use histograms to show the flake-to-flake variability of these results and we fit them with a normalized Gaussian distribution function. Moreover, we plot the corresponding two-dimensional normalized Gaussian distribution function in a 2D gray colormap, in which the density of datapoints is associated with the colorbar, set as inset. The obtained Young's modulus values along the *a*- and *c*-axes directions are $E_{\text{a-axis}} = 44 \pm 8$ GPa and $E_{\text{c-axis}} = 86 \pm 15$ GPa, respectively. Interestingly the anisotropy ratio ($E_{\text{c-axis}}/E_{\text{a-axis}}$) is ~2 among the largest value reported in the literature for anisotropic 2D materials: black phosphorus ~2.7,[17] for orpiment ($As_2S_3$) is ~1.7.[75] Our DFT calculations predict even higher Young's modulus values (91.9 GPa and 216.9 GPa for the a- and c-axis respectively) in relatively good agreement with the values reported for bulk-like $MoO_3$ (~200 nm thick) crystals through Brillouin scattering. Note that the presence of defects in the synthesized $MoO_3$ samples, like the oxygen vacancies found in the STEM-EDS analysis, could explain the lower Young's modulus values obtained in our experiments.

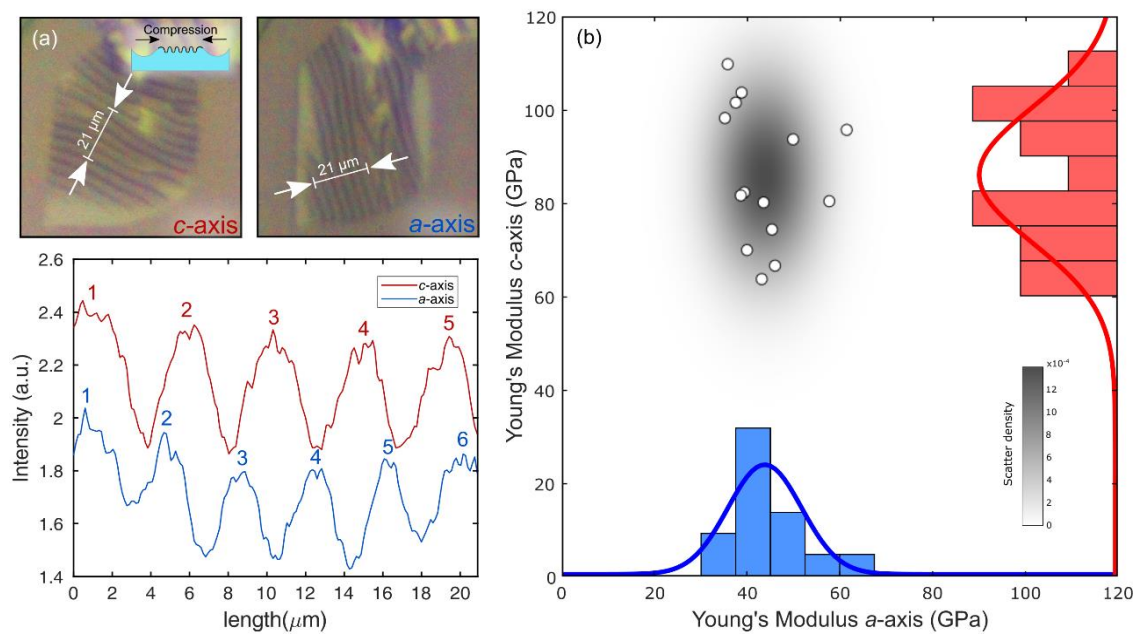

**Fig. 4** (a) Optical images of a MoO$_3$ flake on Gel-Film substrate applying compression along *a* and *c* axes (top images), inset: process of ripples formation; determination of the wavelength of periodic ripples (bottom figure). (b) Scatter density plot of Young's modulus values along the two directions obtained from 14 different samples, white circles with black edges, and fitted with a multivariate Gaussian normal distribution function, in gray. Inset: Histogram plots of Young's modulus along the a-axis (blue) and the c-axis (red), fitted with a Gaussian distribution.

**Conclusions**

In summary, we combined scanning transmission electron microscopy, Raman spectroscopy, micro-reflectance and buckling metrology to probe the anisotropic optical and mechanical properties of exfoliated MoO$_3$ flakes. We found that the flakes show a strong birefringence, i.e. their refractive index depends on the alignment between the polarization of the incident light and the crystalline axis. The difference between the refractive index for light polarized along the *c*- and *a*-axes reaches ~0.1. This large value, together with the fact that MoO$_3$ is transparent in the visible range, makes this material a good candidate for future polarization optics applications. Regarding the mechanical properties, we found that the experimental Young's modulus along the *a*- and *c*-axes directions are $E_{\text{a-axis}} = 44 \pm 8$ GPa and $E_{\text{c-axis}} = 86 \pm 15$ GPa, respectively, yielding an anisotropy ratio of ~2, only surpassed by black phosphorus. Our results agree with ab-initio calculations of optical and elastic properties of MoO$_3$. The anisotropy in the refractive index and the Young's modulus have strong impact in many optical and mechanical properties and thus, we believe that our work can be used as the foundation of further works studying the intriguing anisotropic properties of MoO$_3$.

**Methods**

**Growth and Deposition**

We have based our present grown procedure on a modification of the hot plate growth method developed by Molina-Mendoza et. al[56].

Once the growth of the crystals is finished, the $MoO_3$ flakes are firstly exfoliated onto a polydimethylsiloxane (PDMS) (Gel-Film WF x4 6.0mil, by Gelpak®) and then transferred onto a 297 nm $SiO_2$/Si substrate using a deterministic transfer method[76,77].

**Scanning Transmission Electron Microscopy**

For the scanning transmission electron microscope characterization, we used an aberration-corrected JEOL JEM-ARM 200cF electron microscope operated at 80 kV, equipped with a cold field emission gun and an Oxford Instruments EDS spectrometer.

**Optical Microscopy and Spectroscopy**

Optical microscopy images were acquired using a Motic BA310 MET-T microscope equipped with a 50× 0.55 NA objective and an AMScope MU1803 CMOS Camera. Reflection spectra were collected from a spot of ~1.5–2 μm diameter with a Thorlabs CCS200/M fiber-coupled spectrometer (Thorlabs Inc., Newton, New Jersey, United States). More details about the micro-reflectance setup can be found in Reference [78].

**Raman Spectroscopy**

Raman characterization of $MoO_3$ flakes on 297 nm $SiO_2$/Si substrates were carried out with a confocal Raman microscopy system (MonoVista CRS+ from Spectroscopy & Imaging GmbH) using a 532 nm excitation laser with an incident power of 1.234 mW and a 100× objective with the integration time of 20 s.

**Numerical methods**

We have analyzed the optical and elastic properties of $MoO_3$ using the YAMBO and Quantum Espresso suite of programs[79–81]. For the electronic structure calculation we have used the generalized gradient approximation in the PBE parametrization for the exchange-correlation energy with a plane wave cut-off of 60 Ry[82]. The electronic structure calculations are performed on a Monkhorst-Pack grid of 8x8x8 points. We have chosen norm-conserving pseudo potentials in the SG15 database[83].

For the elastic properties, we have used the thermo_pw code from the Quantum Espresso suite[84]. The elastic properties agree when we used either norm-converting or ultra-soft pseudo potentials.

**Data Availability**

Data presented in this study are available on request from the authors.

**Acknowledgements**

This project has received funding from the European Research Council (ERC) under the European Union's Horizon 2020 research and innovation programme (grant agreement n° 755655, ERC-StG 2017 project 2D-TOPSENSE), the European Commission, under the Graphene Flagship (Core 3, grant number 881603), the Spanish Ministry of Economy, Industry and Competitiveness through the grant MAT2017-87134-C2-2-R. R.F. acknowledges the support from the Spanish Ministry of Economy, Industry and Competitiveness (MINECO) through a Juan de la Cierva-formación fellowship 2017 FJCI-2017-32919. SP acknowledges the fellowship PRE2018-084818. R. D'A. acknowledges financial support from the grant Grupos Consolidados UPV/EHU del Gobierno Vasco (Grant No. IT1249-19), the support of the MICINN through the grant "SelectDFT" (Grant No. FIS2016-79464-P) and travel support from the MINECO grant "TowTherm" (Grant No. MINECOG17/A01). G.S.-S. acknowledges financial support from Spanish MICIU RTI2018-099054-J-I00 and MICINN IJC2018-038164-I. Electron microscopy observations were carried out at the Centro Nacional de Microscopia Electronica, CNME-UCM.

**Author Contributions**

S.P. fabricated the MoO$_3$ flake samples and performed the optical microscopy, Raman spectroscopy and buckling metrology experiments. G.S-S., performed the electron microscopy experiments. R.F., C.M. and A.C.-G. supervised and designed the experiments. S.P., C.M. and A.C.-G. drafted the firsts version of the manuscript. R. D'A. developed the theoretical calculations. All authors contributed to writing the final version of the manuscript.

**Competing interests**

The authors declare no competing interests.

**Supplementary Information**

# In-plane anisotropic optical and mechanical properties of two-dimensional MoO$_3$

Sergio Puebla [1], Roberto D'Agosta [2,3], Gabriel Sanchez-Santolino[4], Riccardo Frisenda [1], Carmen Munuera [1] and Andres Castellanos-Gomez [1,*]

[1] Materials Science Factory. Instituto de Ciencia de Materiales de Madrid (ICMM-CSIC), Madrid, E-28049, Spain.

[2] Nano-Bio Spectroscopy Group and European Theoretical Spectroscopy Facility (ETSF), Departamento Polímeros y Materiales Avanzados: Física, Química y Tecnología, Universidad del País Vasco UPV/EHU, Avenida Tolosa 72, E-20018 San Sebastián, Spain.

[3] IKERBASQUE, Basque Foundation for Science, Plaza Euskadi 5, E-48009 Bilbao, Spain

[4] GFMC, Departamento de Física de Materiales & Instituto Pluridisciplinar, Universidad Complutense de Madrid, 28040 Madrid, Spain

\* Correspondence: andres.castellanos@csic.es

In order to explain the Angle-Resolved Polarized Raman Spectroscopy technique we provide the following Figure S1a, in which is shown a scheme of the experimental set up used to obtain the data shown in Fig. 1. In Figure S1b we show the laboratory coordinates, as dotted arrows, and the sample coordinates, as continuous arrows.

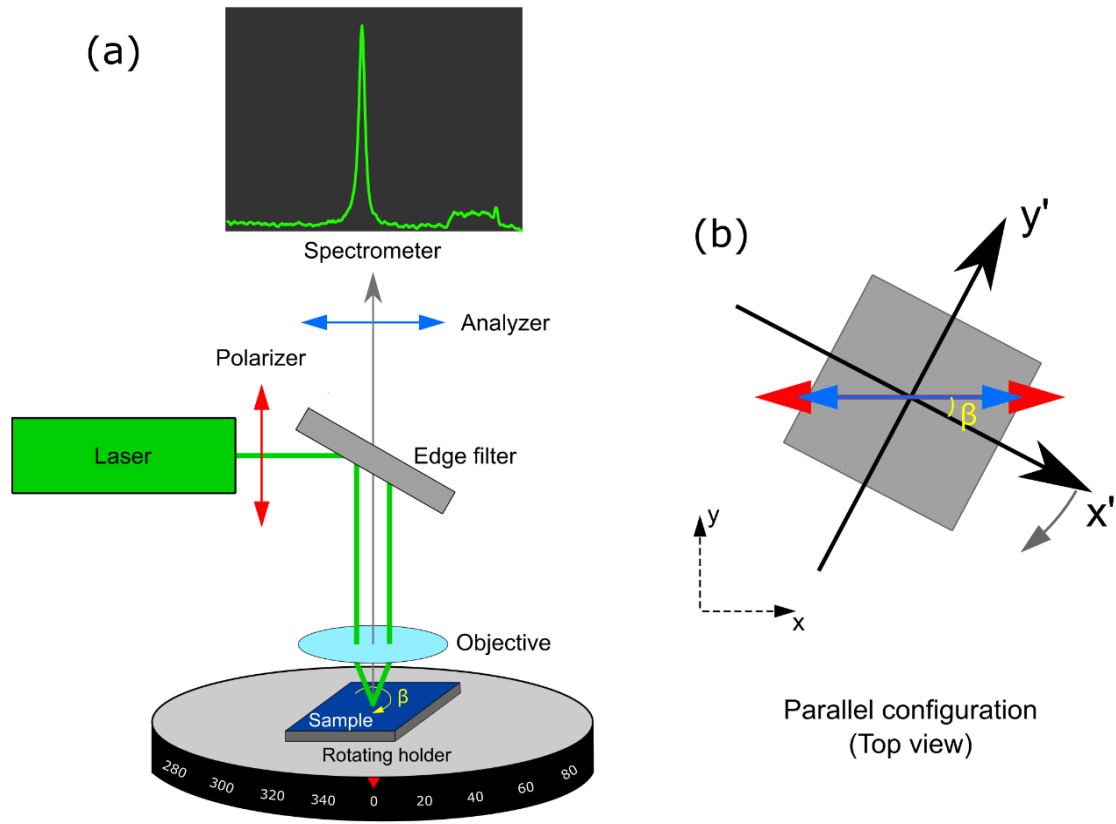

**Fig. S1** (a) Schematic diagram of polarization configuration setup used for angle-resolved polarized Raman scattering spectroscopy, in which the incident light pass through a linear polarizer and reaching the sample, which is placed onto a rotating holder. The Raman scattered light is passed through the analyzer before entering the spectrometer. (b) Top view of the configuration, the laser and analyzer are set in parallel configuration (both are oriented in the same direction) and the grey curved arrow indicates the rotation of the sample with its axis with an angle β respect to the laboratory coordinates. Laboratory coordinates (xyz) are represented by black dotted arrows, and crystal coordinates are represented by black continuous arrows.

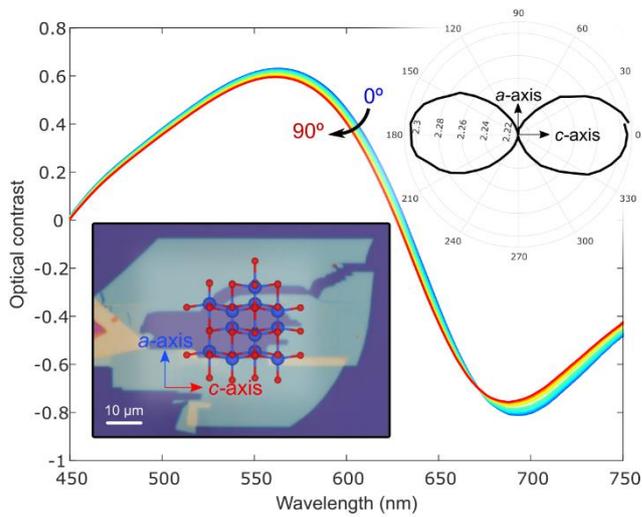

**Fig. S2** Optical contrast of MoO$_3$ flake of thickness 24 nm, studied in the main text Fig. 4. Insets: Polar plot of refractive index (top right) and picture of the flake (bottom left), showing its axes and its respective crystal structure.

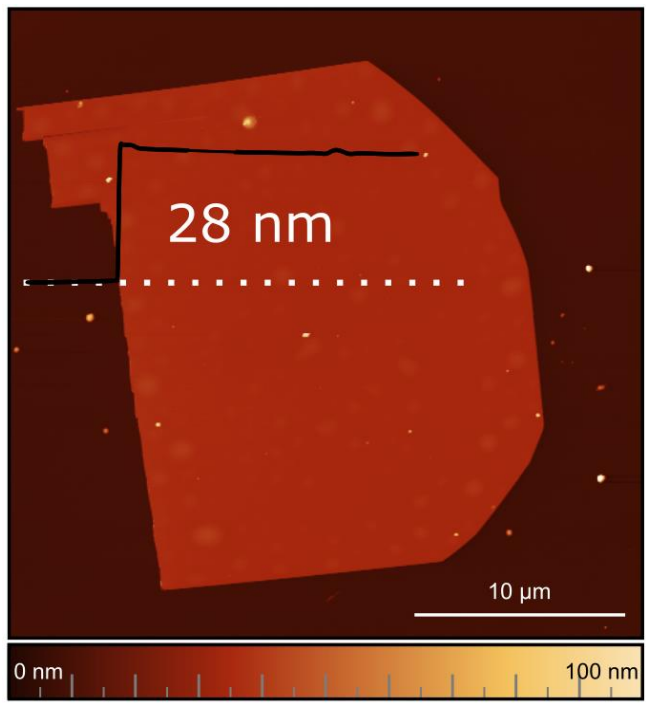

**Fig. S3**. AFM measurement of MoO$_3$ flake shown in Fig. 2b of main text. Scale bar is 10 μm.

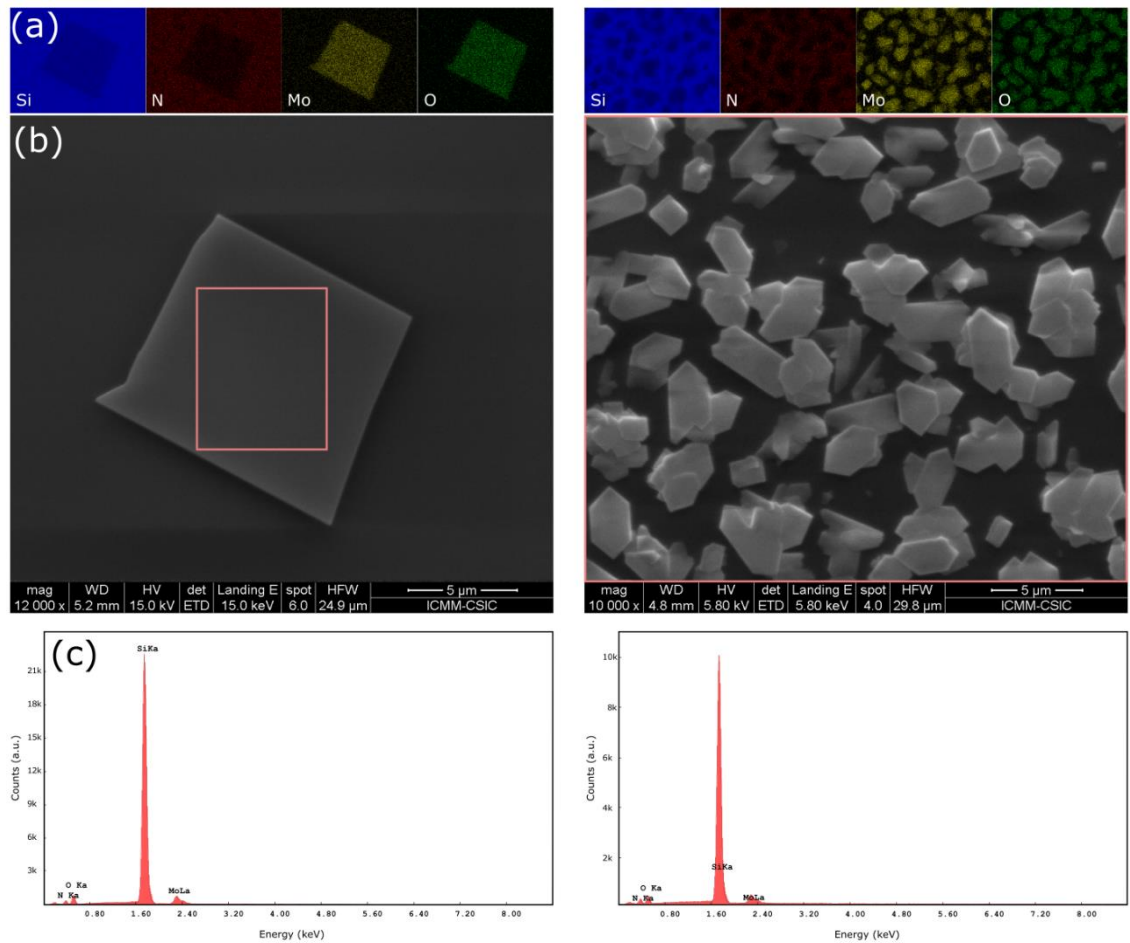

**Fig. S4** (a) Energy dispersive x-ray spectroscopy (EDS) -SEM map of a MoO$_3$ flake onto a Si substrate with 200 nm of Si$_3$N$_4$ capping layer, showing each element contribution. (b) SEM image showing the area in red, in which EDS measurement has been carried out. (c) EDS measurement showing N, O, Si and Mo peaks. Left part of the figure corresponds to a flake deposited on the substrate with all-dry deterministic placement and the right part corresponds to flakes directly grown on a Si$_3$N$_4$/Si substrate.

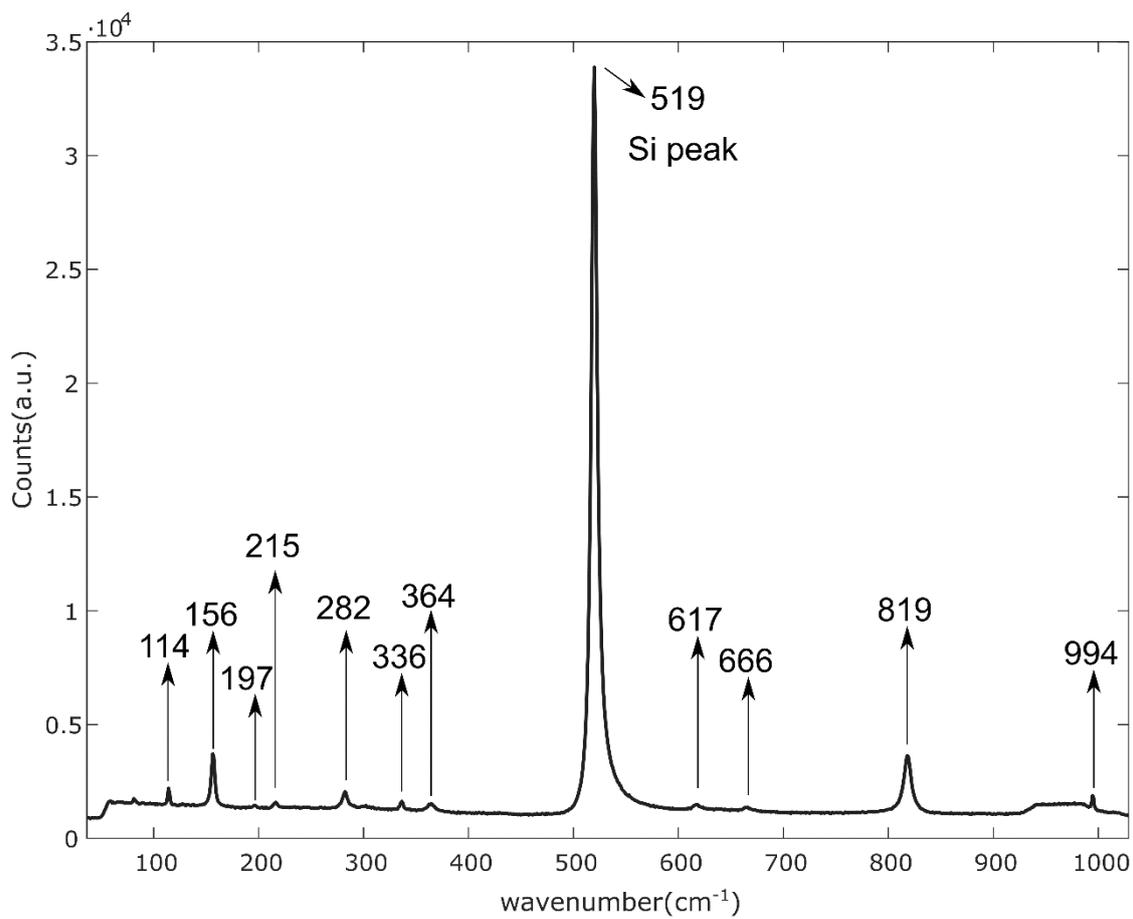

**Fig. S5.** Full Raman spectra of $MoO_3$ sample, pointed peaks belong to typical $MoO_3$ Raman spectrum,[1–3] except the one of 519 cm$^{-1}$ belonging to Si. No signal arising from $MoO_2$ has been observed.

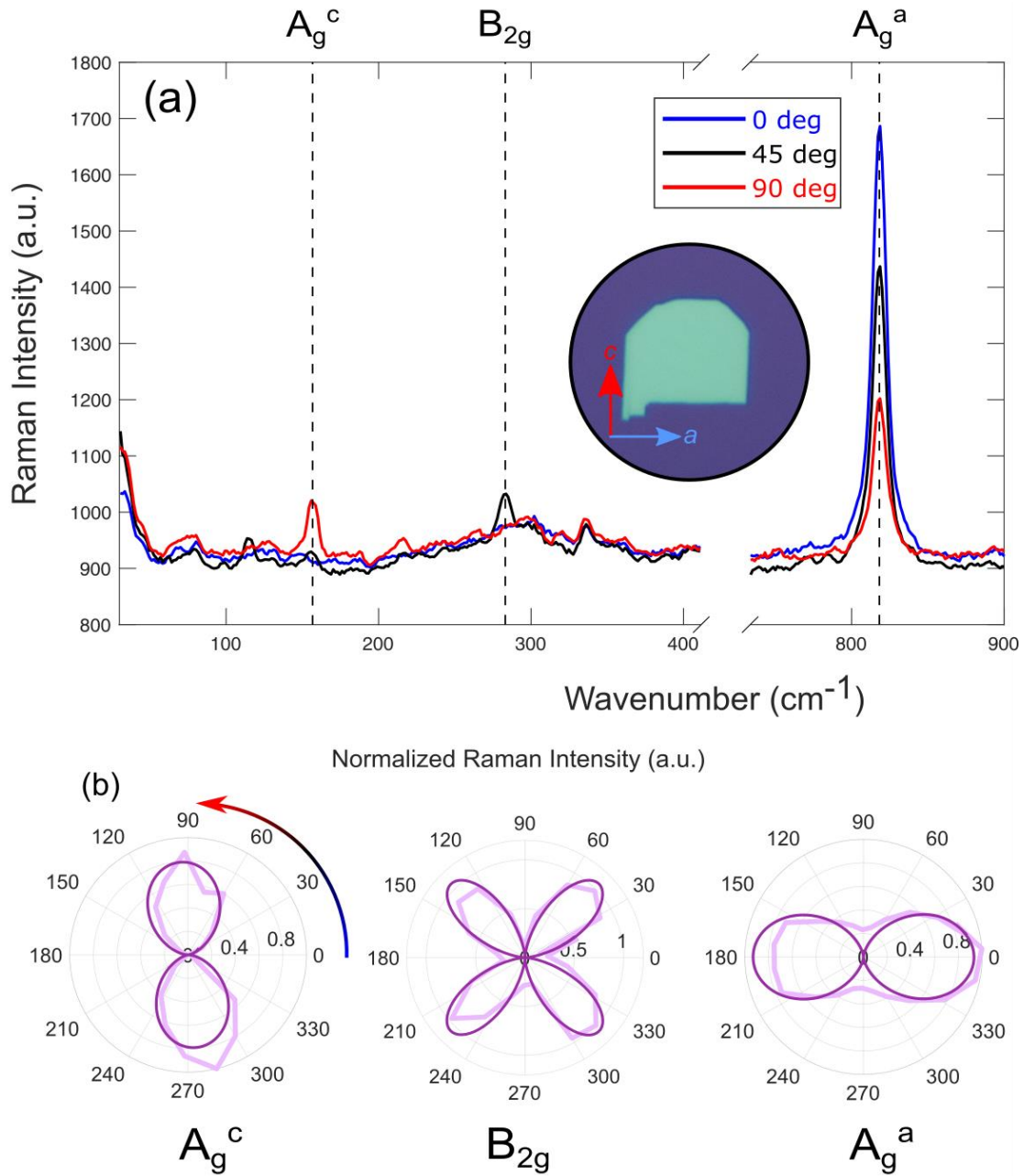

**Fig. S6.** (a) Raman spectra at 0º (blue), 45º (black) and 90º (red) angles with respect to the horizontal axis. $A_g^c$, $B_{2g}$ and $A_g^a$ Raman modes are highlighted with vertical dashed lines. Inset: Microscopy image of the sample placed in the initial position (0º) and the *c* and *a* axes, determined from the angle-resolved Raman measurements, are shown. (b) Polar plots, in light purple; and fittings, in dark purple, of angle-resolved normalized Raman intensities of the $A_g^c$ (left), $B_{2g}$ (middle) and $A_g^a$ (right) modes. These measurements have been done using a 457nm laser wavelength and the same method as the one in Fig. 1. Note, $MoO_3$ Raman peaks have no dependence on the wavelength of laser used.

*Theoretical Analysis*

We have investigated the elastic response of the MoO$_3$ orthorhombic phase with crystal lattice $a$ = 3.969 Å, $b$ = 14.425 Å, $c$ = 3.761 Å. Using DFT and the Quantum-Espresso package (thermo_pw) and norm-conserving pseudo potentials, we have calculated the elastic matrix (all elements in units of GPa)

$$C = \begin{pmatrix} 110.50 & 27.32 & 34.60 & 0 & 0 & 0 \\ 27.32 & 42.05 & 32.22 & 0 & 0 & 0 \\ 34.60 & 32.22 & 243.57 & 0 & 0 & 0 \\ 0 & 0 & 0 & 38.80 & 0 & 0 \\ 0 & 0 & 0 & 0 & 59.47 & 0 \\ 0 & 0 & 0 & 0 & 0 & 30.93 \end{pmatrix}$$

The stiffness matrix, $S$ (1/GPa), the inverse of $C$, is,

$$S = \begin{pmatrix} 0.01088128 & -0.00654958 & -0.00067912 & 0 & 0 & 0 \\ -0.0065496 & 0.03040684 & -0.00309197 & 0 & 0 & 0 \\ -0.00067909 & -0.00309199 & 0.00461108 & 0 & 0 & 0 \\ 0 & 0 & 0 & 0.02577902 & 0 & 0 \\ 0 & 0 & 0 & 0 & 0.01681517 & 0 \\ 0 & 0 & 0 & 0 & 0 & 0.03232566 \end{pmatrix}$$

From this we can read the elastic coefficients along the principal axis:

**Young moduli**

$E_a$=91.9 GPA, $E_c$=216.87 GPa,

while the **Poisson's ratios** are

$\nu_{ca} = 0.147$, $\nu_{ac} = 0.06$

Similar results are obtained with the ultra-soft pseudo potential.

Following the Reuss and Voigt approximations for polycrystalline materials we get

Voigt approximation:

Bulk modulus $B =$ 64.932164 GPa

Young modulus $E =$ 111.302287 GPa

Shear modulus $G =$ 45.829385 GPa

Poisson Ratio $\nu =$ 0.21431

Reuss approximation:

Bulk modulus $B =$ 39.591641 GPa

Young modulus $E =$ 77.938993 GPa

Shear modulus $G =$ 33.253128 GPa

Poisson Ratio $\nu =$ 0.17190

Voigt-Reuss-Hill average of the two approximations:

Bulk modulus $B =$ 52.261903 GPa

Young modulus $E =$ 94.620640 GPa

Shear modulus $G =$ 39.541257 GPa

Poisson Ratio $\nu =$ 0.19648

**Supporting Information references:**